\documentclass[useAMS,usenatbib,twocolumn,usegraphicx]{mn2e}

\usepackage{times}

\newcommand{\gsim}{\, \raisebox{-0.8ex}{$\stackrel{\textstyle >}{\sim}$ }}

\newcommand{\beq}{\begin{equation}}
\newcommand{\eeq}{\end{equation}}
\newcommand{\beqar}{\begin{eqnarray}}
\newcommand{\eeqar}{\end{eqnarray}}

\title[Constraints on the Magnetic Field Geometry of Magnetars ] {Constraints on the Magnetic Field Geometry of Magnetars }

\author[H. Sotani, A. Colaiuda, \& K. D. Kokkotas]
{H. Sotani$^1$\thanks{E-mail:sotani@astro.auth.gr}, A.
Colaiuda$^2$\thanks{E-mail:colaiuda@tat.physik.uni-tuebingen.de},
and K. D. Kokkotas$^{1,2}$\thanks{E-mail:kostas.kokkotas@uni-tuebingen.de}
\\
  $^1$Department of Physics, Aristotle University of Thessaloniki,
  Thessaloniki 54124, Greece \\
  $^2$Theoretical Astrophysics, University of T\"{u}bingen, Auf der Morgenstelle 10,
  72076, T\"{u}bingen, Germany}
\begin{document}


\maketitle


\begin{abstract}
We study the effect of the magnetic field geometry on the oscillation spectra of strongly magnetized stars.
We construct a configuration of magnetic field where a toroidal component is added to the standard poloidal one. 
We consider a star with a type I superconductor core so that  both components of the magnetic 
field are expelled from the core and confined in the crust. Our results show that 
the  toroidal contribution does not influence significantly the torsional oscillations of
the crust. On the contrary, the confinement of the magnetic field in the crust drastically affects
on the torsional oscillation spectrum. 
Comparison with estimations for the magnetic field strength,
from observations, exclude the possibility that magnetars will have a magnetic field solely confined
in the crust i.e. our results suggest that the magnetic field in whatever geometry has to permeate
the whole star.
 
\end{abstract}


\section{Introduction}
\label{sec:Intro}

It is known that magnetic fields  play important role in
the astrophysical phenomena, such as supernovae, gamma-ray bursts,
galaxy jet, and so on. Recently, there is a growing consensus in
explaining Soft Gamma Repeaters (SGRs) via the magnetar model
\citep{DUN1992}. Magnetars are believed to be neutron stars with strong magnetic field
which is responsible for the observed flare activity.
Three giant flares,  SGR 0526-66, SGR 1900+14, and SGR 1806-20,
have been detected so far.
The peak luminosities are in the range of $\simeq
10^{44}-10^{46}$ erg/s. This huge amounts of energy can be explained by
the presence of a strong magnetic field whose strength is
estimated to be larger than $4\times 10^{14}$ Gauss for SGR 1900+14
\citep{SGR1900} and of the order of $ 8\times 10^{14}$ Gauss for SGR
1806-20 \citep{SGR1806}. In addition to the initial short and hard peak, these three events show a
decaying tail lasting hundreds of seconds. In this late part careful analysis revealed the existence 
of characteristic Quasi Periodic Oscillations (QPOs); see
\cite{SW2006} for references. The oscillation frequencies in
these QPOs are in the range of a few tenths of Hz to kHz
and thought to be associated with crust torsional modes of
magnetars. These signals might be the first evidence of a direct
detection of crust oscillations of neutron stars:  a first attempt
to explain these frequencies by the crust torsional oscillations
has been done by using a model with dipole magnetic fields and
realistic EoS (see \cite{SA2007} and \cite{SKS2007}) and
this attempt was partially successful.
On the other hand \cite{GSA2006} suggested an alternative scenario
to explain the observational frequencies via  Alfv\'{e}n
oscillations while \cite{Levin2006} was the first to realize that the Alfv\'{e}n
oscillations of magnetars could be a continuum. More recently
\cite{Levin2007} pointed out that the edges or turning point of the
continuum can be long-lived QPOs. In a recent paper,
\cite{SKS2007b} showed that the Alfv\'{e}n oscillations
of magnetars could be a continuum which can explain  the lower ones of the observed frequencies.

However, it is quite difficult to construct realistic models
for neutron star magnetic fields and/or to understand  
the details of their dynamics. In fact, the observational
data suggest that  the surface temperature of neutron stars
with a strong magnetic field is not uniform (e.g.,
\cite{SHHM2005,HTV2006}). This observational evidence can be
explained, for example,  by the presence of magnetic fields in the crust region
with a strong toroidal components or a more complicated
structure than the poloidal one \citep{GKP2006,PG2007}. Such
crustal magnetic fields can be generated  if the core is a  type I
superconductor. In this  case, the magnetic fields are expelled
from the core due to the Meissner effect after about $10^4$ years
\footnote{Note that there is a possibility that the flux expulsion is more local, 
and will not lead to the entire core being void of flux
(e.g., \cite{SC1998}).}.

Recently, the analysis of the spectrum of timing noise for SGR
1806-20 and SGR 1900+14 has supported the idea
that the core region can be a type II superconductor \citep{ACT2004}.
For the type II superconductor, even if an initial uniform magnetic field is
present in the core, it will be split in thin magnetic fluxes
tubes (\textit{fluxoids}), which are parallel to the initial
magnetic field. These fluxoids interact with the neutron vortices
and are driven by these vortices outside the core, towards the
crust on a timescale of $10^8-10^9$ years \citep{Ruderman}. In any case, it is too
difficult to discriminate between all these models by using only
the past observational data.  This is due to uncertainties 
associated with the
chemical composition of the interior and the crust of neutron
stars as well as about the interior magnetic field configuration
and its influence on the properties of neutron star.

The aim of this paper is to calculate the axisymmetric crust torsional modes of magnetars
with both poloidal and toroidal magnetic field, where  both the components are confined in the crust.
Then our numerical results show that this magnetic configuration cannot explain 
the actual observational data of SGRs.
In this paper,
we adopt the unit of $c=G=1$, where $c$ and $G$ denote the speed
of light and the gravitational constant, respectively, and the
metric signature is $(-,+,+,+)$.

\section{Equilibrium Configuration}
\label{sec:II}

The deformation from  spherical symmetry due to the magnetic
pressure would be small, because the
magnetic energy is considerably smaller than the gravitational
energy. So in this paper we neglect the effect of deformations
induced by the magnetic field as well as the rotational
deformations since  the known magnetars rotate slowly. For these
reasons, the equilibrium model of neutron stars can be described
by a spherically symmetric solutions of the Tolman-Oppenheimer-Volkov
(TOV) equations. The line
element is given then by
\begin{equation}
 ds^2 = -e^{2\Phi}dt^2 + e^{2\Lambda}dr^2 + r^2(d\theta^2 + \sin^2\theta d\phi^2),
\end{equation}
and the 4-velocity is expressed as $u^{\mu}=(e^{-\Phi},0,0,0)$. In
the following, we assume an ideal MHD approximation and we
consider an axisymmetric magnetic field  generated by a 4-current
$J^\mu$. In order to determine the distribution of magnetic
fields we need two basic relations: the Maxwell equations and the equation of motions that
can be obtained by projecting the conservation of the  energy-momentum tensor $T^{\mu \nu}$ on
to the hypersurface normal to $u^\mu$ (see equation (11) in \cite{SKS2007}). 
These two equations are:
\begin{eqnarray}
 F^{\mu\nu}_{\ \ ;\nu} &=& 4\pi J^{\mu}, \label{Maxwell} \\
 (\epsilon + p)u^{\nu}u_{\mu;\nu} + p_{,\mu} + u_{\mu} u^{\nu} p_{,\nu} &=& F_{\mu\nu}J^{\nu}, \label{Eular}
\end{eqnarray}
where $F_{\mu\nu}$ is the Faraday tensor  linked to the
electromagnetic 4-potential $A_{\mu}$ by
$F_{\mu\nu}=A_{\nu,\mu}-A_{\mu,\nu}$. With the appropriate gauge
condition, we can set $A_{\mu}$  as
\begin{equation}
 A_{\mu} = (0,A_r,0,A_{\phi}).
\end{equation}
 With $J^r$ and $J^{\theta}$ obtained from the equation
(\ref{Maxwell}), the equation (\ref{Eular}) for $\mu\equiv\phi$ can be
written as
\begin{equation}
 \tilde{A}_{r,\theta}A_{\phi,r} - \tilde{A}_{r,r}A_{\phi,\theta}=0,
\end{equation}
where $\tilde{A}_r$ is defined as $\tilde{A}_r \equiv
e^{\Phi-\Lambda}A_{r,\theta}\sin\theta$. Thus $\tilde{A}_r$ 
depends on $A_{\phi}$ and we can set  $\tilde{A}_r=\zeta
A_{\phi}$ (where $\zeta$ is a constant, see below for its meaning) since $\tilde{A}_{r}$ is of the same order as $A_{\phi}$
with respect to the magnetic field strength. If we expand $A_{\phi}$ as
\begin{equation}
 A_{\phi} = a_{\ell_M}(r)\sin\theta \partial_{\theta}P_{\ell_M}(\cos\theta),
\end{equation}
then, the 4-potential $A_{\mu}$ has the form
\begin{equation}
 A_{\mu} = (0,\zeta e^{-\Phi+\Lambda}a_{\ell_M}P_{\ell_M},0,a_{\ell_M}\sin\theta \partial_{\theta}P_{\ell_M}).
\end{equation}
In the following, we will consider only dipole fields i.e. $\ell_M=1$.

In order to obtain an expression for the 4-current $J^\mu$, we can
get additional equations from the equation (\ref{Eular}) by setting
$\mu\equiv r,\theta$:
\begin{equation}
 \chi_{,r} = A_{\phi,r}\tilde{J}_{\phi} \ \ \ \mbox{and}\ \ \
 \chi_{,\theta} = A_{\phi,\theta} \tilde{J}_{\phi}, \label{chi}
\end{equation}
where $\chi$ is a function of $r$ and $\theta$ and
\begin{eqnarray}
 \tilde{J}_{\phi} &\equiv& \frac{1}{r^2\sin^2\theta (\epsilon + p)}
     \left[J_{\phi} - \frac{\zeta^2}{4\pi}e^{-2\Phi}A_{\phi}\right]
     \label{jtilde}, \\
 \chi_{,r} &\equiv& C_s^2(\ln n)' + \Phi' \label{chider}.
\end{eqnarray}
Note that $C_s$ is the sound speed, while in order to derive the
above equations we have used the first law of thermodynamics, i.e,
$d\epsilon = (\epsilon + p)dn/n$, where $n$ is the  number density
of particles. Using the integrability condition
$\chi_{,r\theta}=\chi_{,\theta r}$ together with equation
(\ref{chi}), we obtain $A_{\phi,r}\tilde{J}_{\phi,\theta} -
A_{\phi,\theta}\tilde{J}_{\phi,r} = 0$. Thus we can see that
$\tilde{J}_{\phi}$ is also function of $A_{\phi}$ and it is
possible to rewrite $\tilde{J}_{\phi}$ as $\tilde{J}_{\phi}=-f_0 -
f_1 A_{\phi}$, where $f_0$ and $f_1$ are some constants. Comparing
this last expression with equation (\ref{jtilde}), we get the
following expression for $J_{\phi}$:
\begin{equation}
 J_{\phi} = \frac{\zeta^2}{4\pi}e^{-2\Phi} A_{\phi} - (f_0 + f_1 A_{\phi})(\epsilon + p)r^2\sin^2\theta.
     \label{Jphi}
\end{equation}
For simplicity we neglect the contribution of $f_1$
\footnote{Actually the term of $f_1$ can contribute in the
equation (\ref{GS}) as the term of order $\ell_M=3$.
However we use $f_1=0$ since the contributions from terms with $\ell_M>2$ are negligible. Additionally,
the omission of the $f_1$ term leads to the condition adopted in \cite{KOK1999} in the limit of $\zeta\to 0$.}.

Setting $\mu\equiv\phi$ in  Maxwell equations (\ref{Maxwell}) and
by making use of equation (\ref{Jphi}) we derive the following equation for
$a_1$
\begin{equation}
 e^{-2\Lambda}\frac{d^2a_1}{dr^2} + (\Phi' - \Lambda')e^{-2\Lambda}\frac{da_1}{dr}
     + \left(\zeta^2e^{-2\Phi}-\frac{2}{r^2}\right)a_1 = -4\pi j_1, \label{GS}
\end{equation}
where $j_1 = f_0 r^2 (\epsilon + p)$. For the exterior region, if
we assume only  dipole magnetic fields, i.e., $\zeta=0$,
there is an analytic solution for $a_1^{\rm{(ex)}}$ that is:
%
\begin{equation}
 a_1^{\rm{(ex)}} = -\frac{3\mu_br^2}{8M^3}\left[\ln\left(1-\frac{2M}{r}\right) + \frac{2M}{r}
     + \frac{2M^2}{r^2}\right],
     \label{out_a1}
\end{equation}
where $\mu_b$ is the magnetic dipole moment observed at infinity.
The interior solution for $a_1$ should be determined by solving
numerically equation (\ref{GS}) and requiring that 
$a_1$ and ${a_1}'$ are continuous across the stellar surface. Hence, the components of
the vector $H_{\mu}\equiv B_{\mu}/\sqrt{4\pi}$, are given by
\begin{eqnarray}
 H_r        &=& \frac{e^{\Lambda}a_1}{\sqrt{\pi}r^2}\cos\theta, \\
 H_{\theta} &=& -\frac{e^{-\Lambda}{a_1}'}{\sqrt{4\pi}}\sin\theta, \\
 H_{\phi}   &=& -\frac{\zeta e^{-\Phi}a_1}{\sqrt{4\pi}}\sin\theta.
\end{eqnarray}
>From these expressions, we can see that the parameter $\zeta$ is a
constant associated only with the toroidal component of the magnetic field.
Additionally the value of $\zeta$ is chosen in the range where the
function of $a_1$ has no node inside the star. Therefore there exists
a maximum value of $\zeta$ named as $\zeta_{max}$.
Note that we can produce magnetar models with larger value of $\zeta$ than $\zeta_{max}$.
But for these models, the structure of the magnetic field becomes more complex
since the sign of current is not constant in the star,
and this could lead to possible instability of the magnetic fields.
In other words, when the sign of current changes,
even if the star remains stable the magnetic field structure might undergo
a catastrophic instability which will restructure the field.
So we examine only the magnetar models with $\zeta \le \zeta_{max}$.

\begin{figure}
\begin{center}
\includegraphics[height=3.5cm]{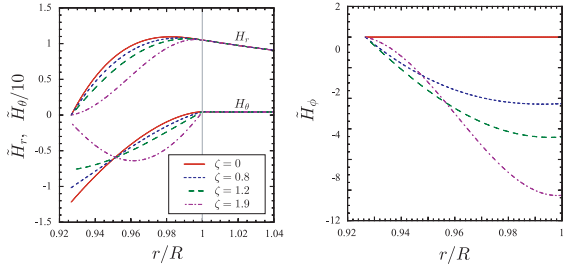}
\caption{
The three components, ${\tilde H}_r\equiv R^3H_r/\mu_b$ (for
$\theta=0$), ${\tilde H}_\theta\equiv R^2H_\theta/\mu_b$ (for
$\theta=\pi/2$) and ${\tilde H}_\phi\equiv R^2H_\phi/\mu_b$ (for
$\theta=\pi/2$) of the magnetic field for a stellar model with
$1.4M_{\odot}$. The different
lines correspond to different values of the constant $\zeta$ with
$\zeta_{max}=1.9$. Note also that ${\tilde H}_\theta$ is scaled 
by a factor ten.
}
\label{fig:field-(ii)}
\end{center}
\end{figure}

In this paper we consider a stellar model
with a superfluid core (type I superconductor) so that the magnetic field
is expelled from the core by the Meissner effect. As a consequence, the magnetic fields
are confined only in the crust region.
In order to construct the model, we have to impose
the condition that the magnetic fields are vanishing inside the core, i.e., we
require that $a_1$ should be zero at the basis of crust. This
condition can be written as $a_1 \simeq \alpha_c (r-R_c)$,
%
%
%
where $\alpha_c$ is a constant and $R_c$ is the radius at the basis of the crust.
Note that at $r=R_c$ both $H_r$ and $H_\phi$ vanish while $H_\theta$ is not zero.
In Figure \ref{fig:field-(ii)}, we plot the components of magnetic field,
where $\tilde{H}_r$, $\tilde{H}_\theta$,
and $\tilde{H}_\phi$ are defined as $\tilde{H}_r\equiv R^3H_r/\mu_b$,
$\tilde{H}_\theta\equiv R^2H_\theta/\mu_b$, and
$\tilde{H}_\phi\equiv R^2H_\phi/\mu_b$. 
The actual details of the used stellar models will be described in the next section. 
It is worth noticing, from these figures, that the $\theta$ components of
magnetic fields are not smooth at the stellar surface,
because we assume an exterior magnetic field with
only a poloidal component, which means $\zeta=0$ for $r\geq R$ i.e. no
toroidal component outside the star.
This choice produces a current on the star surface. The presence of a surface current 
is found also in many simulations dealing with the evolution and the stability of the magnetic field
(see \cite{BR2006}).
Howevere we do not consider the existence of the surface current in this paper,
since the surface current dose not affect the behavior of the torsional oscillations
within our analysis.

%

\section{Numerical results}
\label{sec:III}

In this paper,  we adopted 
polytropic equations of state (EoS) described by the following relations
\begin{equation}
 p    = Kn_0m_b \left(\frac{n}{n_0}\right)^\Gamma \ \mbox{and}\ \ \
 \rho = nm_b + \frac{p}{\Gamma-1},
\end{equation}
where $m_b=1.66\times 10^{-24}$ g and $n_0=0.1$ fm$^{-3}$, and we
have set $\Gamma = 2.46$ and $K=0.00936$ by fitting the tabulated
data for EoS A \citep{EOS_A}. With this equation of state and with the fixed
value of the density at $r=R_c$, $\rho=2.4\times 10^{14}$
g/cm$^3$, the maximum mass  is $M_{max}\simeq 1.654M_{\odot}$.
Here  we study two representative stellar
models; one with $M=1.4M_{\odot}$ and the other is the one with the maximum allowed mass for this EoS
i.e. the one with $M=1.654M_{\odot}$. For these two stellar models, the  radius,
the relative crust thickness, and the
compactness are $R=10.35$ km, $\Delta r/R=7.34$ \%, and
$M/R=0.200$ for the model with $M=1.4M_{\odot}$ while the model with $M=1.654M_{\odot}$ has
$R=8.95$ km, $\Delta r/R=4.15$ \%, and $M/R=0.273$.
Finally, for simplicity the shear modulus in the
crust, $\mu$, is assumed to be given by a relation of the
formula $\mu=v_s^2 \rho$, where $v_s$ is the
speed of shear waves with a typical value
$v_s\simeq  10^8$ cm/s \citep{ST1983}.
In general the frequencies of torsional oscillations depend on the EoS in the crust region
since the shear modulus also depend on the EoS.
However,  in \cite{SKS2007} we have used two different realistic EoS for the crust and our results
show that
the freuqencies of fundamental torsional oscillations are almost independent of them.
Anyway in this paper, as  first approximation, we use the above simple formula for shear modulus.

We study  axial perturbations in the  Cowling approximation of the previous stellar models.
We focus only on these type of oscillations of the crust (torsional) since they do not induce
density perturbations, which means that significantly lower energy is needed for their excitation.
On the other hand the absence of density perturbations leads to minimal metric perturbations
which justifies our choice in using the Cowling approximation. Under these assumptions the
perturbation equations are reduced to equation
(51) in \cite{SKS2007}. By studying  this equation one can observe that,
if the background magnetic fields are axisymmetric,
the axisymmetric torsional oscillations are independent from the component $H_\phi$
of the magnetic field. Thus we can
calculate the eigen-frequencies by using the perturbation
equations (69) and (70) in \cite{SKS2007} with exactly the same boundary conditions,
such as ${\cal Y}_{,r}=0$ at the basis of crust ($r=R_c$) and at the stellar surface ($r=R$),
where ${\cal Y}(r)$ describes the angular displacement of the stellar material.
However, we have to replace $j_1$ in equation (70) in \cite{SKS2007}
by $\tilde{j}_1=j_1+\zeta^2e^{-2\Phi}a_1/(4\pi)$, because for the
elimination of the term of ${a_1}''$ we have used  equation (\ref{GS}).

\begin{figure}
\begin{center}
\includegraphics[height=4cm]{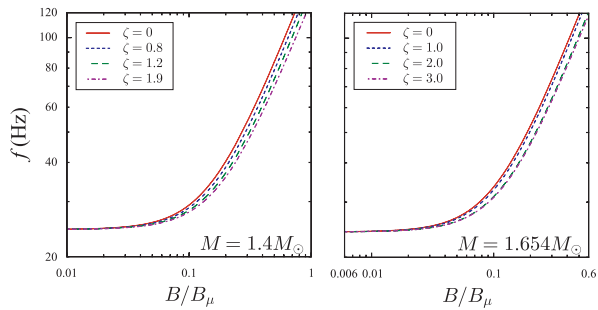}
\caption{
With $M=1.4M_{\odot}$ and $1.654M_{\odot}$,
the frequencies of the fundamental torsional modes of $\ell=2$
as functions of $B/B_{\mu}$, where $B_\mu = 4\times 10^{15}$ Gauss.
The different lines  correspond to  different values of $\zeta$,
where $\zeta_{max}\sim 1.9$ for $M=1.4M_{\odot}$ and $\zeta_{max}\sim 3.0$ for $M=1.654M_{\odot}$.
}
\label{fig:type(ii)}
\end{center}
\end{figure}

The numerical results are shown in Figure \ref{fig:type(ii)},
where the fundamental $\ell=2$ torsional modes are plotted
as function of the normalized magnetic field $B/B_\mu$
($B_\mu=4\times 10^{15}$ Gauss).
>From these figures, it is
clear that the frequencies depend weakly on the presence of the
toroidal magnetic fields, even if the strength of the
toroidal field increases i.e. the value of $\zeta$ becomes
larger. 
This weak dependence 
can be understood if we recall that the magnetic field lines of the toroidal component are
parallel to the direction of fluid motion. 
But the most important result is that,
although it seems that qualitatively the effect of
the magnetic field is similar to the results in \cite{SKS2007}, 
there is a significant quantitative difference:
the influence of the magnetic field on the axisymmetric torsional frequencies
becomes apparent for considerably lower strengths of the magnetic field,
i.e. already for $B/B_\mu \approx 0.1$.
Notice that the results in \cite{SKS2007} show that the effect of the magnetic field
can be seen for $B/B_\mu\gsim 1$.
It is natural that when the magnetic field is confined in the crust,
magnetic field becomes considerably stronger than the one that permeates the whole star 
for the same strength of the exterior magnetic field. 
Actually, since the magnetic field is confined in the crust which has 
thickness $\Delta r\sim 0.03-0.12 R\approx 0.5-1.5$km, the density of
magnetic field lines becomes higher.
Also if we compare the results for the two stellar models under discussion
we observe that thinner is the crust, then earlier the effect of the magnetic field
becomes pronounced. For example, for the model with $M=1.654M_{\odot}$ ($\Delta r/R=4.15$ \%)
the effect of magnetic field is apparent for $B/B_\mu \approx 0.07$.
For the other model ($M=1.4M_{\odot}$ and $\Delta r/R=7.34$ \%)
the effect of the magnetic field becomes pronounced later, i.e., $B/B_\mu \approx 0.1$.

\begin{figure}
\begin{center}
\includegraphics[height=5.5cm]{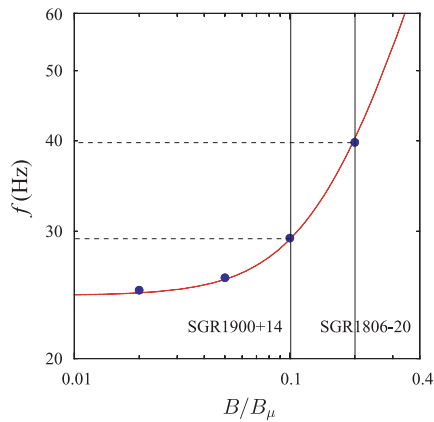}
\caption{
The torsional mode frequencies as functions of the magnetic field strength
for $\zeta=0$.
The continuous line shows the results from the 1-D truncated mode calculation
on which we omitted  the coupling to $\ell\pm 2$ equations
while the thick dots correspond to results from 2D numerical evolutions of the perturbation equations.
In this figure  two vertical lines has been drawn  corresponding to the estimated magnetic field strengths
for the two SGRs that is  $B\approx 4\times 10^{14}$ Gauss for SGR 1900+14 and
$B\approx 8\times 10^{14}$ Gauss for SGR 1806-20.
}
\label{fig:comparison}
\end{center}
\end{figure}

The above results are derived
by using the same procedure as in \cite{SKS2007},
i.e. we decompose the system into spherical harmonics and
truncate the ($\ell \pm 2$) terms. This truncation could introduce small errors in the results,
thus in order to check the validity of our results we have used an alternative technique based on
2D numerical evolutions of the perturbation equation describing the dynamics of the oscillating crust.
The 2D method used here is similar to the one described in \cite{SKS2007b} and the evolution equations
were structurally similar although the various background terms have been modified to include terms
related to the shear modulus, $\mu$.
The 2D evolution was constrained only in the crust and in the first quadrant
i.e. $0\le \theta \le \pi/2$ and the following boundary conditions have been imposed:
(a) ${\cal Y}_{,r}=0$ at $r=R_c$ and $R$,
(b) ${\cal Y}_{,\theta} =0$ at $\theta=0$, and (c) ${\cal Y}=0$ at $\theta = \pi/2$.
The results of the two approaches for $\zeta=0$ are plotted in in Figure \ref{fig:comparison}.
The continuous line corresponds to results derived by using the truncated equations
and treating the problem as a boundary value one while
the thick dots correspond to results produced by the 2D numerical evolution.
The agreement between the two approaches is apparent,
which suggests that for this type of problems the truncation of the ($\ell \pm 2$)
equations does not affect at all the results.

Furthermore, we performed similar calculations for stellar models with different realistic EoS,
but the behavior of frequencies, as function of a magnetic field, is similar to that for
the EoS A that we use here.

%

%
%

\section{Constraints on the Magnetic Field Geometry}
\label{sec:IV}

We have studied the torsional oscillations for the magnetic field distribution
where the magnetic fields has been ``expelled" from the superfluid core and confined in the crust.
We have also assumed that the magnetic field has both poloidal and toroidal components.
The study shows that the presence of the toroidal field affects marginally the oscillation frequencies
of the torsional modes, which can be explained because the perturbations propagate along the toroidal field lines.
The toroidal field might play a more significant role in the case of non-axisymmetric perturbations or
if we include the coupling to polar perturbations.

The main results of this work is summarized in Figure \ref{fig:comparison}.
Our calculations show that the effect of magnetic field on the torsional modes,
when it is confined only in the crust, 
can be observed for magnetic field strengths below $10^{14}$ Gauss.
Moreover, this observation has a significant effect on understanding the geometry of
the magnetic field of magnetars. The reason is the following.
With this magnetic field configuration there are no additional oscillational frequencies
such as Alfv\'{e}n oscillations (see \cite{SKS2007b}),
because the magnetic field does not exist in the core.
Thus the $\ell=2$ torsional oscillation is lowest oscillational frequency.
On the other hand the estimated magnetic field strengths are of
the order of  a few times $10^{14}$ Gauss \citep{SGR1900,SGR1806} or even higher \citep{NYY2007}.
This suggest that if the magnetic field was constrained to the crust
the lowest possible frequency for the signals from SGR 1806-20 is about 40 Hz.
Then, this magnetic field distribution
can not explain the observed frequencies 18, 26 and 29 Hz.
So this type of magnetic field geometry cannot be present
in the observed magnetars and most probably it has to be excluded in general for the case of
strongly magnetized neutron stars. Alternative scenarios which might include coupling with
the polar modes should be excluded since crustal perturbations of polar type are typically of
considerably higher frequencies \citep{Strohmayer1991,VKS2007} and cannot explain all three frequencies.

\section*{Acknowledgments}

We are grateful to N. Stergioulas for helpful comments. This work was supported
in part by the Marie Curie Incoming International
Fellowships (MIF1-CT-2005-021979), the GSRT via the Pythagoras II program,
and the German Foundation for Research (DFG) via the SFB/TR7 grant.
AC is acknowledging the support of V. Ferrari and L. Gualtieri
in constructing magnetar models with toroidal fields.


\appendix
\section[]{Poloidal and toroidal fields permeating the star  }
\label{sec:A1}

\begin{figure}
\begin{center}
\includegraphics[height=7.5cm]{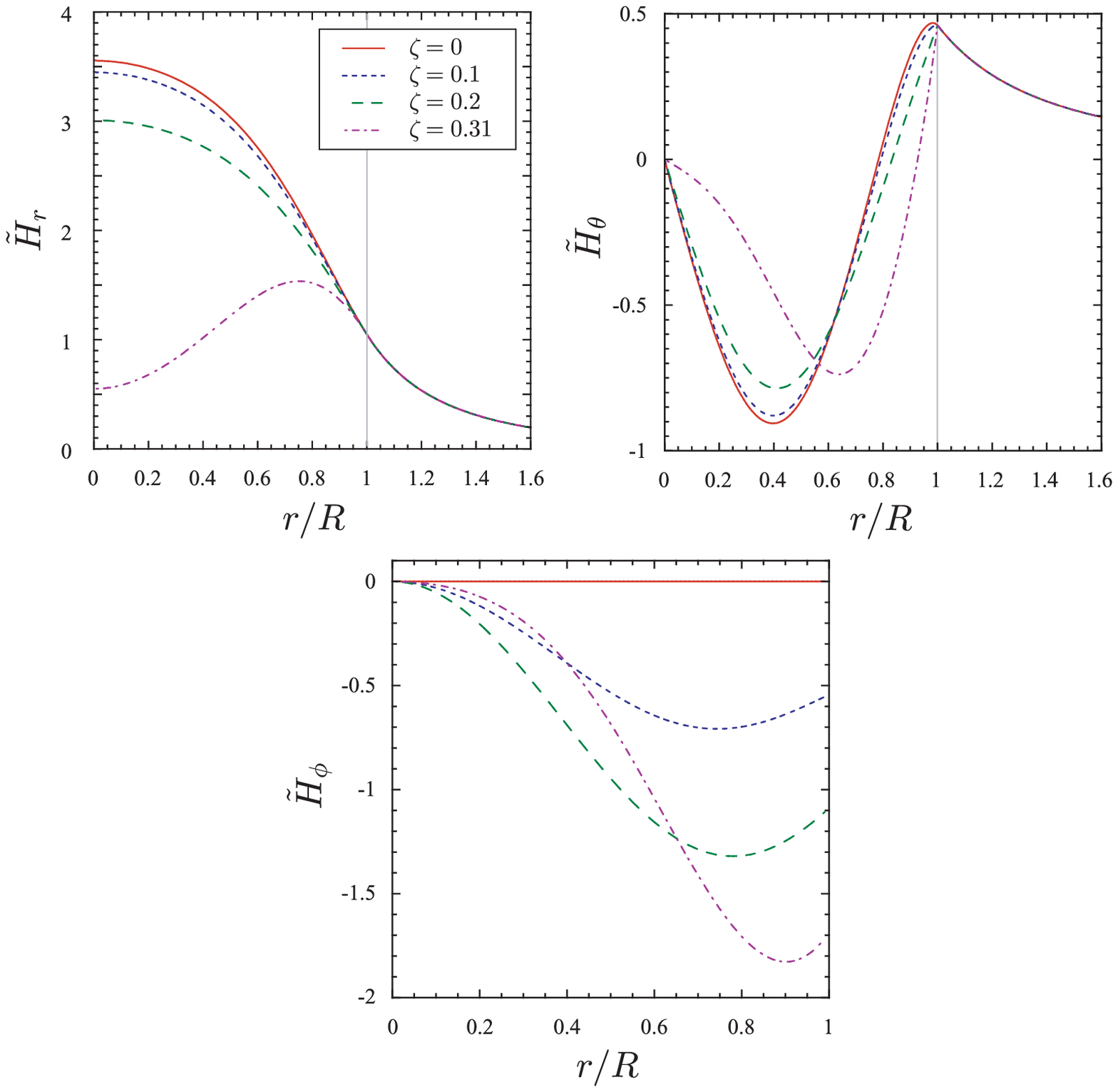}
\caption{
The three components, ${\tilde H}_r\equiv R^3H_r/\mu_b$ (for
$\theta=0$), ${\tilde H}_\theta\equiv R^2H_\theta/\mu_b$ (for
$\theta=\pi/2$) and ${\tilde H}_\phi\equiv R^2H_\phi/\mu_b$ (for
$\theta=\pi/2$) of the magnetic field for a stellar model with
$1.4M_{\odot}$.
 The different
lines correspond to different values of the constant $\zeta$ with
$\zeta_{max}=0.31$.
}
\label{fig:field-(i)}
\end{center}
\end{figure}

In this appendix we study the effect on the torsional mode spectrum of the toroidal component
added on the poloidal magnetic field configuration studied in \cite{SKS2007}. In this case the
magnetic field permeates the whole star but we study only the oscillations of
the crustal part of the star. Although this is somehow a truncation of the full problem,
it may be a good approximation in studying the effect of the toroidal component
(see the next paragraph  for an argument in favour of this). 
Following \cite{SKS2007},  the equilibrium configuration of the magnetic field is 
constructed by imposing the regularity condition,  $a_1\simeq \alpha_c r^2$,
at the stellar center, and by requiring the continuity 
of $a_1$ with the analytic exterior solution (\ref{out_a1}). The contribution of the 
toroidal field is set to zero
for $r \geq R$. The components of the magnetic field are plotted in 
Figure \ref{fig:field-(i)} in a similar fashion 
as in Figure \ref{fig:field-(ii)}.

%

\begin{figure}
\begin{center}
\includegraphics[height=4cm]{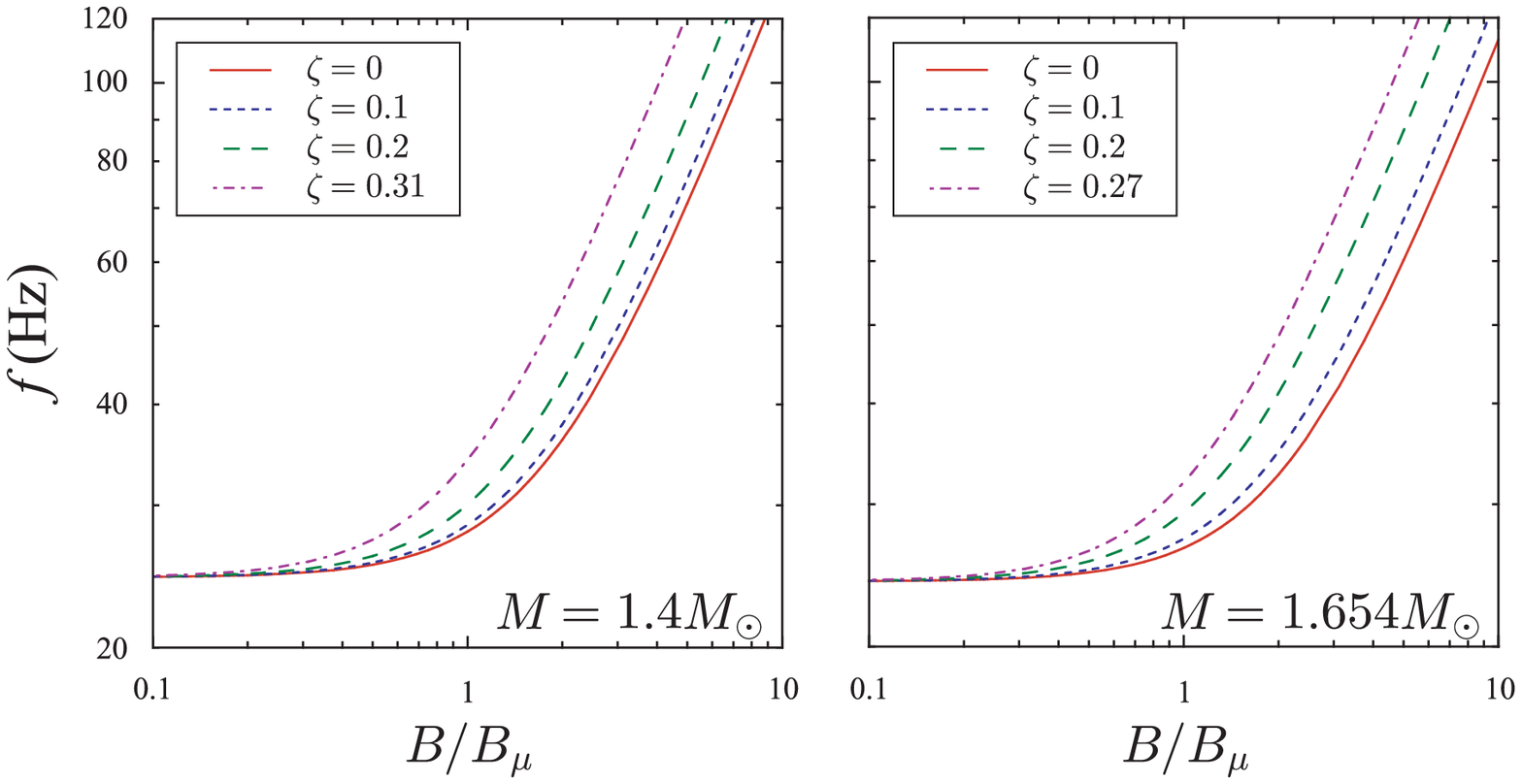}
\caption{
The frequencies of the fundamental torsional modes with $\ell=2$
as functions of $B/B_{\mu}$ 
with $M=1.4M_{\odot}$ and $1.654M_{\odot}$. 
The different lines correspond to different values of $\zeta$,
where $\zeta_{max}\sim 0.31$ for $M=1.4M_{\odot}$ 
and $\zeta_{max}\sim 0.27$ for $M=1.654M_{\odot}$.
}
\label{fig:type(i)}
\end{center}
\end{figure}
%

The torsional oscillation frequencies are calculated following the same procedure as in \cite{SKS2007},
i.e., 1-D truncated mode calculation using the same boundary conditions, i.e., ${\cal Y}_{,r}=0$
at $r=R_c$ and $r=R$.
Actually, the boundary condition at $r=R_c$ is not quite correct because in this way we ignore
the coupling with the fluid core \citep{SKSV2007}.
Thus the proper boundary condition is
 $H^rH_r {\cal Y}^{(-)}_{,r} = \left[\mu + H^rH_r\right]{\cal Y}^{(+)}_{,r}$,
which can be written as  
${\cal Y}^{(+)}_{,r}/{\cal Y}^{(-)}_{,r}\approx \beta^2/(\beta^2+1)$,
where $\beta=B/B_\mu$.
>From this relation it is evident that for magnetic fields with strength around $B/B_{\mu} \simeq 0.1-0.3$
the coupling with the fluid core is not so important for the study of pure crust torsional oscillations. 
The corrections to our approximate boundary condition  arise only  for $B/B_{\mu} \gsim 1$.
Thus approximately we adopt the simple boundary condition ${\cal Y}_{,r}=0$ at the basis
of crust, which will be used to demonstrate the effect of the toroidal component on the spectrum of
torsional oscillations.

In Figure \ref{fig:type(i)} we show the frequencies of the fundamental
$\ell=2$ torsional modes  as function of the normalized magnetic field strengh, $B/B_\mu$. 
As mentioned in \cite{SKS2007}, the effect of magnetic field on
torsional modes becomes important for $B\sim B_\mu$. This is expected since
for $B>B_\mu$ the Alfv\'{e}n speed, $u_A\equiv B/(4\pi\rho)^{1/2}$,
becomes larger than the shear speed, $u_s\equiv (\mu/\rho)^{1/2}$.
>From these figures,  it is clear that the presence of the toroidal magnetic field affects  weakly the
frequencies of the torsional oscillations. 

Concluding, the presence of a toroidal magnetic field component
does not affect significantly the spectrum of the crust oscillations neither
in the case when the poloidal field permeates the star nor when it is confined in the crust.
This suggests that although we are able to exclude the presence of crust confined magnetic fields
in magnetars, most probably we will not be able via the QPOs observations to identify the presence
of a toroidal component in the magnetic field.

\end{document}